\documentstyle[12pt,epsfig,rotating]{article}
\title{
\vspace*{-15mm}
{\normalsize \begin{tabular}[t]{ll}
    HERA-B   & \hspace{-0.5em}98--113\\
Software     & \hspace{-0.5em}98--018\end{tabular}
  \hfill Revised October 25, 2005 \\}
\vspace*{15mm}
{\bf Optimized Integration of the Equations  \newline 
~~~~~~~~~~~~~~of Motion of a Particle  
in the \newline HERA-B Magnet } 
}

\author{
       {\it Alexander Spiridonov}\\
        DESY  Zeuthen~/~ITEP Moscow\\
      E-mail: spiridon@ifh.de } 
\date{July 16, 1998}

\begin{document}
\maketitle
%\tableofcontents
\maketitle
\begin{abstract}
In this note we present a flexible 
%computation 
approach to perform the propagation of track parameters and their derivatives
in an inhomogeneous magnetic field, keeping the computational effort small.
%effort at low level. 
We discuss also a Kalman filter implementation
using this optimized computation of the derivatives.
%The revised version 
%includes also details of Kalman filter implementation
%adopted for the magnet tracking to optimize the computation.
\end{abstract}

\section {Introduction}

Dedicated hadronic B factory experiments as HERA-B are designed as
forward spectrometers, in adjustment to the huge Lorentz boost under
which the B decay particles are produced. The HERA-B tracking
concept~\cite{strategy} relies on the propagation of track candidates
which have been found in the pattern tracker in the field-free area
upstream through the spectrometer magnet. The {\it concurrent track
evolution} strategy~\cite{conc} which employs the Kalman filter
technique is used to cope with the large track densities and still
give a high track finding efficiency. Practical implementation of this
concept requires a fast and precise procedure to transport both the
track parameters and its covariance matrix estimate, in spite of the 
inhomogeneity of the magnetic field.
First solution based on a fifth-order Runge-Kutta method
have been shown in~\cite{Oest}. 

This note presents the mathematical
basis and the program implementation of an approach 
which was developed to achieve a further gain in
speed, and at the same time warrant sufficient accuracy for an optimal
operation of the track finding process. This was achieved by providing
a set of methods which are optimized for different transport ranges,
and by testing the validity of each approximation directly within the
track finding application~\cite{ranger,magnet}.

The Kalman filter technique is used very often in  
high energy physics experiments.
We include Sec.~\ref{applic}
in the revised version of the note
to discuss the optimized implementation
of the Kalman filter for magnet tracking, which
reduces strongly the amount of computation.
It was not described in detail in \cite{ranger,magnet}. 

\section{Equations of motion and solution}

In the following we will use a coordinate system in which 
the $z$ axis points along the proton beam, the $x$ axis is directed 
normal to it in the horizontal plane, pointing towards the inside of
the HERA ring, and the $y$ axis is oriented upwards such that $x$, $y$ 
and $z$ form a right-handed system. This system is identical to the
{\em Arte} coordinate system defined in the appendix of ~\cite{rfit}.

%As track parameters, the following 
%choice will be used which is typical for fixed target experiments
%with relatively small transverse momenta with respect to the longitudinal
%axis: 
The following choice of track parameters is suited 
for fixed target experiments 
with relatively small transverse momenta. 
%with respect to the longitudinal
%will be used for track parameters.
%will be for track parameters.
A particle with momentum $\vec{p}$, charge $Q_e$
and coordinates $\vec{x}$ is described 
by the state vector at a reference $z$ coordinate:
%\begin{equation} 
%\label{x} 
%\tilde{x}^T(z)=(x,\; y,\; t_{x},\; t_{y},\; q)\,, 
%\end{equation} 
\begin{equation}
\label{x}
\tilde{x}(z) = \left (
\begin{array}{c}
x \\ y \\ t_{x} \\ t_{y} \\ q \end{array} \right)\,,
\end{equation}
where $x$ and $y$ are the transverse coordinates, $t_{x}=p_{x}/p_{z}$
and  $t_{y}=p_{y}/p_{z}$ are the track slopes, and
%the ratio 
$q=Q_e/\left|\vec{p}\right|$. 
%is used as momentum parameter
%at the $z$ value of reference. 
%where $x$ and $y$ are the transverse coordinates, 
%$t_{x}=p_{x}/p_{z}$
%and  $t_{y}=p_{y}/p_{z}$ are the track slopes, and
%the ratio $q=Q/\left|\vec{p}\right|$ of the charge $Q$ and the particle 
%momentum $\left|\vec{p}\right|$. 
%\newline Let us use the following  units : 
%\newline 
%{\em    }\hspace{2.in} $\;\;\;\;x,y,z$ in centimeters , 
%\newline  
%\hspace{2.in} $\;\;\;\;p$ in $GeV/c$ , 
%\newline  
%\hspace{2.in} $\;\;\;\;Q_e$ in multiples of the positive elementary 
%charge (dimensionless), 
%\newline  
%\hspace{2.in} $\;\;\;\;\vec{B}$ in kGauss. 
%\newline Let us use the following  units :
%\newline
The parameter $Q_e$ is the particle charge in units 
of the elementary charge. 
%(dimensionless). 
We use the following units: $x,y,z$ are in centimeters,
$p$ is in GeV/c and the magnetic field $B$ in kGauss.

The trajectory of a particle in a static magnetic field
$\vec{B}(\vec{x})$, neglecting stochastic perturbations as
energy loss and multiple scattering,
must satisfy the equations of motion~{~\cite{Bock}}~: 
\begin{equation}
\begin{array}{ll}
\label{dxdz}
dx/dz = & t_{x},  \\
dy/dz = & t_{y}, \\
dt_{x}/dz = & q \cdot \upsilon \cdot A_{x}(t_{x},t_{y},\vec{B}), \\
dt_{y}/dz = & q \cdot \upsilon \cdot A_{y}(t_{x},t_{y},\vec{B}), \\
q= & q_{0}, 
\end{array}  
\end{equation}
where parameter $\upsilon$ is proportional to the velocity of light and is
therefore defined as
\[  \begin{array}{l}
\upsilon = 0.000299792458\,\,(GeV/c)\,kG^{-1}\,cm^{-1} \;\; \\
\end{array}   \]
and the functions $A_{x}$,$A_{y}$ are
\[  \begin{array}{l}
A_{x}=(1+t_{x}^{2}+t_{y}^2)^{\frac{1}{2}} \cdot 
\left[ t_{y}\cdot(t_{x}B_{x}+B_{z})
-(1+t_{x}^{2})B_{y} \right] \;\; ,\\
A_{y}=(1+t_{x}^{2}+t_{y}^2)^{\frac{1}{2}}\cdot 
\left[-t_{x}\cdot(t_{y}B_{y}+B_{z})
+(1+t_{y}^{2})B_{x} \right] \;\; .\\
\end{array}   \]
The initial values at $z=z_0$ are 
\begin{equation}
\label{x0}
\tilde{x}_{0}^T=(x_{0},\; y_{0},\; t_{x0},\; t_{y0},\; q_{0}). 
\end{equation}
Let us assume we should propagate the particle parameters from plane $z_{0}$
to plane $z$.  For solution of the equations (\ref{dxdz}) three different methods 
are used. The choice depends on the distance 
$s=z-z_{o}\;$ between these planes.  

1. {\em $\left|s\right| < 20~{\rm cm}$}: a parabolic expansion of the particle 
trajectory is used

\[  \begin{array}{cl}
x(z)= & x_{0}+t_{x0}\cdot s + \frac{1}{2} \cdot q_{0}\cdot \upsilon \cdot A_{x} \cdot 
s^{2} \;\;,\\    
y(z)= &y_{0}+t_{y0}\cdot s + \frac{1}{2} \cdot q_{0} \cdot \upsilon \cdot A_{y} \cdot
s^{2} \;\;\\
t_{x}(z)= & t_{x0} + q_{0} \cdot \upsilon \cdot A_{x} \cdot s \;\;,\\
t_{y}(z)= & t_{y0} + q_{0} \cdot \upsilon \cdot A_{y} \cdot s \;\;. 

\end{array}    \]

2. {\em$ 20~{\rm cm} \leq \left|s\right| < 60~{\rm cm}$}: 
the classical fourth-order 
Runge-Kutta method ~\cite{RK} is selected to find solution of the
equations (\ref{dxdz}) .

3. {\em $\left|s\right| \geq 60~{\rm cm}$}: a fifth-order Runge-Kutta method
with adaptive step size control ~\cite{RK} is used.

Use of the Kalman filter technique ~\cite{Kalman} for pattern
recognition requires that particle parameters and their covariance 
matrix can be transported to the location of the
next measurement. 
Evaluation of the derivatives of the state vector components with respect
to their initial values (\ref{x0}) is needed to transport the covariance matrix. 
This is achieved by integrating the equations
for derivatives together with the `zero trajectory' (\ref{dxdz}). 

\section{Equations for derivatives}

Let us assume the magnetic field is smooth enough
and field gradients  can be neglected. 
In this case, the equations (\ref{dxdz}) are
invariant with respect to small shifts by $x$ and $y$. This means that
derivatives with respect to initial $x_{0}$, $y_{0}$ are trivial :

\[  \begin{array}{cl}

\partial \tilde{x}^T / \partial x_{0}= &
( 1,\; 0,\; 0,\; 0,\; 0 ) \;\;\;, \\
\partial \tilde{x}^T / \partial y_{0}= &
( 0,\; 1,\; 0,\; 0,\; 0 ) \;\;\;. 
\end{array}    \]
To obtain equations for $\partial \tilde{x} / \partial t_{x0}$, let us
differentiate  equations (\ref{dxdz}) with respect to $t_{x0}$ and change
the order of the derivative operators $\partial  / \partial t_{x0}$ 
and $d/dz$ on the left hand sides :
\begin{equation}
\begin{array}{ll}
\label{dtx}
d/dz(\partial x / \partial t_{x0}) = & \partial t_{x}/\partial t_{x0},  \\
d/dz(\partial y / \partial t_{x0}) = & \partial t_{y}/\partial t_{x0}, \\
d/dz(\partial t_{x} / \partial t_{x0}) = & 
 q_{0} \cdot \upsilon \cdot \left[(\partial A_{x}/\partial t_{x})
( \partial t_{x}/\partial t_{x0}) 
+(\partial A_{x}/\partial t_{y})
( \partial t_{y}/\partial t_{x0}) \right], \\

d/dz(\partial t_{y} / \partial t_{x0}) = & 
 q_{0} \cdot \upsilon \cdot \left[(\partial A_{y}/\partial t_{x})
( \partial t_{x}/\partial t_{x0}) 
+(\partial A_{y}/\partial t_{y})
( \partial t_{y}/\partial t_{x0}) 
\right], \\
\partial q / \partial t_{x0} \ = & 0,
\end{array}  
\end{equation}
where
\[  \begin{array}{ll}
\partial A_{x}/\partial t_{x}= t_{x} \cdot A_{x}/(1+t_{x}^{2}+t_{y}^{2})
+(1+t_{x}^{2}+t_{y}^{2})^{\frac{1}{2}} \cdot
(t_{y} \cdot B_{x} - 2\cdot t_{x}\cdot B_{y}) \; , \\
\partial A_{x}/\partial t_{y}= t_{y} \cdot A_{x}/(1+t_{x}^{2}+t_{y}^{2})
+(1+t_{x}^{2}+t_{y}^{2})^{\frac{1}{2}} \cdot
(t_{x} \cdot B_{x} + B_{z}) \; , \\
\partial A_{y}/\partial t_{x}= t_{x} \cdot A_{y}/(1+t_{x}^{2}+t_{y}^{2})
+(1+t_{x}^{2}+t_{y}^{2})^{\frac{1}{2}} \cdot
(-t_{y} \cdot B_{y} - B_{z}) \; , \\
\partial A_{y}/\partial t_{y}= t_{y} \cdot A_{y}/(1+t_{x}^{2}+t_{y}^{2})
+(1+t_{x}^{2}+t_{y}^{2})^{\frac{1}{2}} \cdot
(-t_{x} \cdot B_{y} + 2\cdot t_{y}\cdot B_{x}) \; . \\
\end{array}  
 \;\;\;\;\;\; 
%(4)   
\]
Initial values for the solution of equations (\ref{dtx}) are :
\begin{equation}
\label{dtx0}
\partial \tilde{x}^T / \partial t_{x0} = 
( 0,\; 0,\; 1,\; 0,\; 0 ).
\end{equation}
The equations for $\partial \tilde{x} / \partial t_{y0}$ are similar
to equations (\ref{dtx}) , but the initial values  are :
\[  \begin{array}{cl}
\partial \tilde{x}^T / \partial t_{y0}= &
( 0,\; 0,\; 0,\; 1,\; 0 ) \;\;. \;\;\;\;\;
%(6)
\\
\end{array}    \]
To obtain equations for $\partial \tilde{x} / \partial q_{0}$, let us
differentiate  the equations (\ref{dxdz}) with respect to $q_{0}$ and change
the order of the derivative operators 
$\partial  / \partial q_{0}$ and $d/dz$
in the left parts : 
\begin{equation}
\begin{array}{ll}
\label{dq}
d/dz(\partial x / \partial q_{0}) = & \partial t_{x}/\partial q_{0} \;,  \\
d/dz(\partial y / \partial q_{0}) = & \partial t_{y}/\partial q_{0} \;, \\
d/dz(\partial t_{x} / \partial q_{0}) = & 
  \upsilon \cdot A_{x} +  \upsilon \cdot q_{0} \cdot
\left[(\partial A_{x}/\partial t_{x})
( \partial t_{x}/\partial q_{0}) 
+(\partial A_{x}/\partial t_{y})
( \partial t_{y}/\partial q_{0}) 
\right]  \;, \\

d/dz(\partial t_{y} / \partial q_{0}) = & 
 \upsilon \cdot A_{y} + \upsilon \cdot q_{0} \cdot
\left[(\partial A_{y}/\partial t_{x})
( \partial t_{x}/\partial q_{0}) 
+(\partial A_{y}/\partial t_{y})
( \partial t_{y}/\partial q_{0}) 
\right]  \;, \\
\partial q / \partial q_{0} = & 1  \; .
\end{array}  
\end{equation}
Initial values for the solution of equations (\ref{dq}) are :
\begin{equation}
\label{dq0}
\partial \tilde{x}^T / \partial q_{0} = 
( 0,\; 0,\; 0,\; 0,\; 1 ).
\end{equation}
%\newpage
In the case of
\[   \begin{array}{cl}
\left|B_{x}/B_{y}\right|, \left|B_{z}/B_{y}\right|, 
\left|t_{x}\right|,\left|t_{y}\right| \ll & 1 \;\;\;\;\;\\
\end{array}    \]
the following relations between $A_{x}$, $A_{y}$ and
their derivatives are valid:

\[  \begin{array}{cl}
\left| \partial A_{x,y}/\partial t_{x,y}\right| & \ll 1 \;\;, \\

\left| \partial A_{x(y)}/\partial t_{x,y}\right| & \ll 
\left|A_{x(y)}\right| \;\;, \\
\left|A_{y}\right|              & \ll \left|A_{x}\right| \;\;.\\
\end{array}  
 \;\;\;\; 
%(9)
   \]
The procedure of evaluating the derivatives is 
simplified when these relations are taken into account. 

{\bf Approximation A}:
The vector  $\partial \tilde{x} / \partial t_{x0}$ can be 
approximated as~: 
\[  \begin{array}{cl} 
 \partial \tilde{x}^T / \partial t_{x0}= &
 (\partial x/\partial t_{x0} ,\;\partial y/\partial t_{x0} ,\; 1,\; 
 \partial t_{y}/\partial t_{x0},\; 0 ) \;\;,\\
\end{array}    \]
where the derivatives are solutions of the system of equations

\[  \begin{array}{ll}
d/dz(\partial x / \partial t_{x0}) = & \partial t_{x}/\partial t_{x0} \;,  \\
d/dz(\partial y / \partial t_{x0}) = & \partial t_{y}/\partial t_{x0} \;, \\
d/dz(\partial t_{x} / \partial t_{x0}) = & 0 \;,\\
d/dz(\partial t_{y} / \partial t_{x0}) = & 
 q_{0} \cdot \upsilon \cdot \left[(\partial A_{y}/\partial t_{x})
( \partial t_{x}/\partial t_{x0}) 
+(\partial A_{y}/\partial t_{y})
( \partial t_{y}/\partial t_{x0}) 
\right]  \;, \\
\partial q / \partial t_{x0} \ = & 0  \; 
\end{array}  
 \;\;\;\;\;\; 
%(11)
   \]
with initial values (\ref{dtx0}).
The solution for the derivatives $\partial \tilde{x} / \partial t_{y0}$
in the same approximation is similar~: 

\[  \begin{array}{cl} 
 \partial \tilde{x}^T / \partial t_{y0}= &
 (\partial x/\partial t_{y0} ,\; \partial y/\partial t_{y0} ,\; 
 \partial t_{x}/\partial t_{y0},\; 1,\;  0 ) \;\;.\\
\end{array}    \]
 The derivatives 
\[  \begin{array}{cl} 
 \partial \tilde{x}^T / \partial q_{0}= &
 (\partial x/\partial q_{0} ,\;\partial y/\partial q_{0} ,\; 
 \partial t_{x}/\partial q_{0},\;\partial t_{y}/\partial q_{0},\; 1 ) \;\;\\
\end{array}    \]
are solutions of the equations (\ref{dq}) 
with the initial values given in (\ref{dq0}). 

{\bf Approximation B}:
 This is the most drastic simplification - the derivatives 
$\partial A_{x,y}/\partial t_{x,y}$  as well as
$A_{y}$ are neglected in the corresponding equations.
The system of equations (\ref{dtx}) is simplified to : 

\[  \begin{array}{ll}
d/dz(\partial x / \partial t_{x0}) = & \partial t_{x}/\partial t_{x0} \;,  \\
d/dz(\partial y / \partial t_{x0}) = & \partial t_{y}/\partial t_{x0} \;, \\
d/dz(\partial t_{x} / \partial t_{x0}) = & 0  \;, \\
d/dz(\partial t_{y} / \partial t_{x0}) = & 0  \;, \\
\partial q / \partial t_{x0} \ = & 0  \; .
\end{array}  
 \;\;\;\;\;\; 
%(10)   
\]
The solution of this system with initial values (\ref{dtx0}) is
\[  \begin{array}{cl}
\partial \tilde{x}^T / \partial t_{x0}= &
( s,\; 0,\; 1,\; 0,\; 0 ) \;\;.  \;\;\;\;\;\\
\end{array}    \]
The solution of similar equations for derivatives 
$\partial /\partial t_{y0}$ is
\[  \begin{array}{cl}
\partial \tilde{x}^T / \partial t_{y0}= &
( 0,\; s,\; 0,\; 1,\; 0 ) \;\;.  \;\;\;\;\;\\
\end{array}  \]
For the vector of derivatives $\partial /\partial q_{0}$ we obtain :
\[  \begin{array}{cl} 
 \partial \tilde{x}^T / \partial q_{0}= &
 (\partial x/\partial q_{0},\; 0,\; 
 \partial t_{x}/\partial q_{0},\; 0,\; 1 ) \;\;,\\
\end{array}    \]
where the derivatives are solutions of the system of equations

%\[  \begin{array}{ll}
%d/dz(\partial x / \partial q_{0}) = & \partial t_{x}/\partial q_{0} \;,  \\
%d/dz(\partial t_{x} / \partial q_{0}) = & 
%  C \cdot A_{x}  \; \\
%\end{array}  
% \;\;\;\;   \]
\begin{equation} 
\begin{array}{l}
\label{deriveq}
d/dz(\partial x / \partial q_{0}) = \partial t_{x}/\partial q_{0} ,  \\
d/dz(\partial t_{x} / \partial q_{0}) =  \upsilon \cdot A_{x},
\end{array}
\end{equation}  
with initial values from (\ref{dq0}).

\section{Application of the Kalman filter technique for the magnet tracking} 
\label{applic}
We use the notations from \cite{Kalman} to describe the 
practical implementation of the Kalman filter technique
for track following in the magnet.
The system state vector (\ref{x}) after inclusion of $k$ measurements
is denoted by $\tilde{x}_k$, and its covariance matrix by $C_k$.
The coordinate measurement corresponding to the hit $k$ 
is denoted by $m_k$. The HERA-B magnet tracking detectors
(drift tubes and Micro-Strip Gaseous Chambers) measured
only one coordinate and $m_k$ is a scalar
%one-dimentional vector 
%(scalar). 
%a number. 
%The 
and its covariance matrix $V_k$ 
%describing the measurement error 
contains only one element.
%is a number also.
The relation between the track parameters $\tilde{x}_k$ and the
expected measurement $m_k$ is described by the projection matrix $H_k$.
The matrix $H_k$ has the structure:
\begin{equation} 
\label{h1}
H^{(1)} = (h_1,\,0,\,\,0,\,\,0,\,\,0)\,,
\end{equation}
for a detector plane measuring only $x$ coordinates
(signal wires of the drift tubes or anode strips of the MSGC
are parallel to the $y$ axis), and
\begin{equation}  
\label{h2} 
H^{(2)} = (h_1,\,h_2,\,\,0,\,\,0,\,\,0)\,, 
\end{equation} 
for stereo planes rotated around the $z$ axis.
The predicted state vector $\tilde{x}^{k-1}_k$ is determined
as the solution of the equations (\ref{dxdz}) with the initial
value $\tilde{x}_{k-1}$.
The covariance matrix of the vector $\tilde{x}^{k-1}_k$ is obtained by
the propagation of the matrix $C_{k-1}$: 
 \begin{equation}   
\label{ck1}  
C^{k-1}_k = F_k C_{k-1} F^T_k + Q_k,  
\end{equation}  
where $Q_k$ denotes the covariance matrix of the process noise
and the transport matrix is
\begin{equation}    
\label{fk}
F_k = \frac {\partial (\tilde{x}^{k-1}_k)}  
{\partial (\tilde{x}_{k-1})}.~~~~~~~~~~~~~~~~  
\end{equation}   
The estimated residual and variance become
\begin{equation} 
\begin{array}{ll} 
\label{rk} 
r^{k-1}_k = m_k - H_k\,\tilde{x}^{k-1}_k\,, & \\
R^{k-1}_k = V_k + H_k\,C^{k-1}_k\,H_k^T\,.  &
\end{array} 
\end{equation} 
The updating of the system state vector after inclusion of
the measurement $k$ is obtained with the filter equations:
\begin{equation}
\label{kxc} 
\begin{array}{ll}
K_k = C^{k-1}_k\,H^T\,(R^{k-1}_k)^{-1}, &  \\
\tilde{x}_k = \tilde{x}^{k-1}_k + K_k\,r^{k-1}_k, &\\
C_k = (1 - K_k\,H_k)\,C^{k-1}_k, &
\end{array}
\end{equation}
with the filtered residuals
\begin{equation}  
\begin{array}{ll}  
\label{rrk}  
r_k = (1 - H_k\,K_k)\,r^{k-1}_k, & \\
R_k = (1 - H_k\,K_k)\,V_k. &
\end{array}  
\end{equation}  
The $\chi^2$ contribution of the filtered point is given by:
%\begin{equation}   
%\begin{array}{ll}   
%\label{chi}   
\[\chi_k^2 = r^T_k\,R^{-1}_k\,r_k.~~~~~~~~~~~~~\] 
%& \\
%\end{array} 
%\end{equation} 
%Let us use notations:
%\[ \{C_{k-1}\}_{ij} = c_{ij} ~~~~{\rm and}~~~~
%\{C^{k-1}_k\}_{ij} = C_{ij}. \]
%------------------------------------------------------
\begin{figure}[h]
\begin{center}
\unitlength1cm
\begin{picture}(15.5,8)
\put( -.80,0.0){\epsfig
%{file=hist_31_1_all.eps,width=8.7cm}}
{file=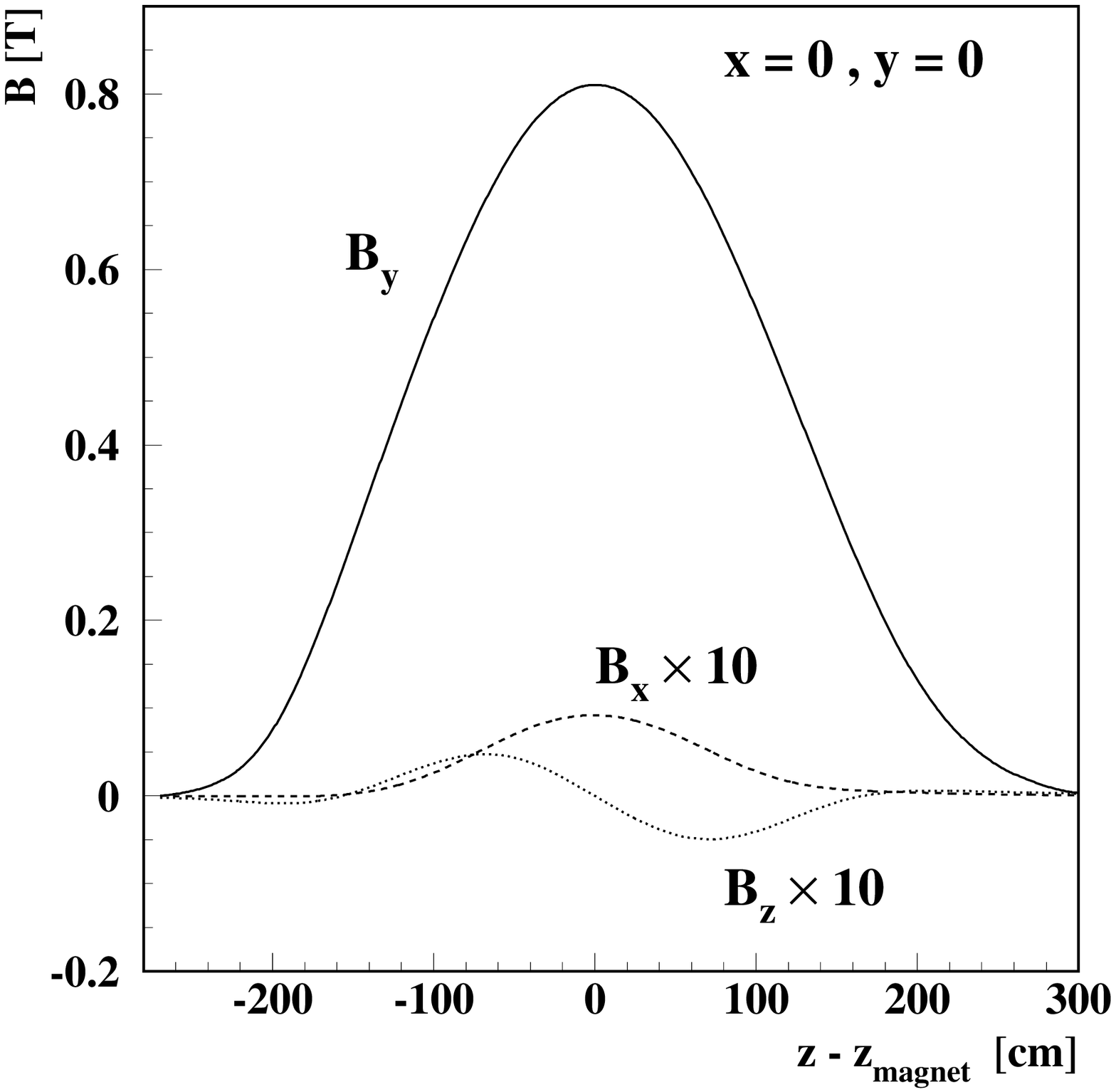,width=8.95cm}} 
\put(6.7,4.9){\makebox(0,0)[t]{\huge a)}}
\put(7.7,0.0){\epsfig
%{file=hist_42_12_all.eps,width=8.7cm}} 
{file=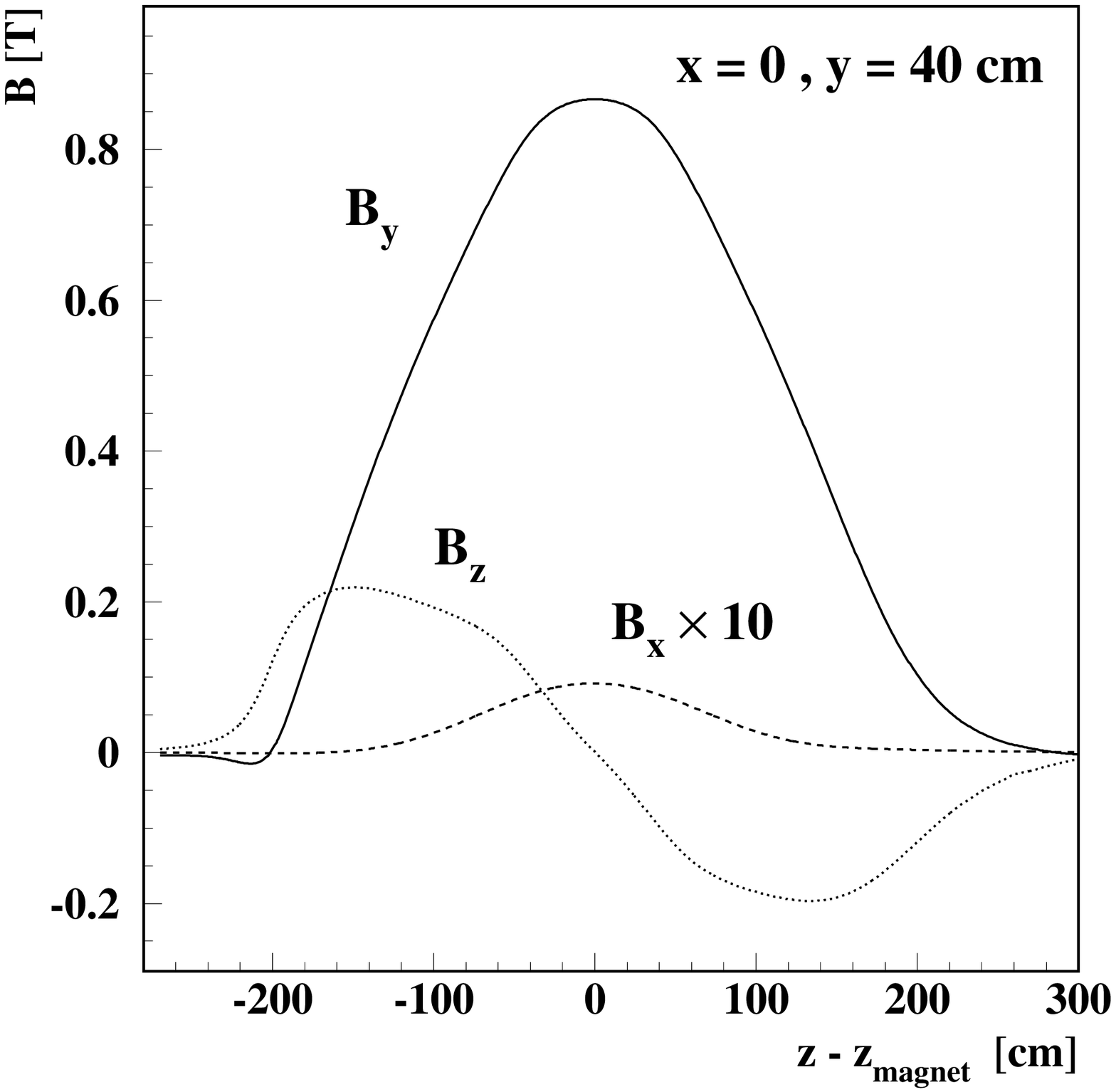,width=8.95cm}}
\put(15.0,4.9){\makebox(0,0)[t]{\huge b)}}
\end{picture}
\caption{\small  
Magnetic field components as a function of the $z$ coordinate
in the magnet center (a) and at a vertical displacement of 40 cm (b).
The $z$ coordinate is given relative to the magnet center at 
$z_{magnet}\,=\,450\,{\rm cm}.$}
\label{field}
\end{center}
\end{figure}
%--------------------------------------------------------

Because of sparse projection matrices (\ref{h1},\ref{h2}),
the calculation in (\ref{rk}--\ref{rrk}) becomes rather simple.
For the matrix $H^{(2)}$ the variance of
estimated residuals is  
\[
\{R^{k-1}_k\}_{11} = \{V_k\}_{11} + 
h_1\,(h_1\,C_{11}\,+h_2\,C_{21})\,+\,
h_2\,(h_1\,C_{12}\,+h_2\,C_{22}), 
\]
i.e. it involves 6 multiplications. Here we use the notation 
$C_{ij} = \{C^{k-1}_k\}_{ij}.$
For the same case, each of the five elements of the gain matrix
is calculated as
\[
 \{K\}_{i1} = k_i = (C_{i1}\,h_1\,+\,C_{i2}\,h_2)\,\,
\{R^{k-1}_k\}^{-1}_{11}.
\]
The calculation of the gain matrix includes 15 multiplications and 1 division. 
The most time consuming operations are the propagation of
the covariance matrix in (\ref{ck1}) and the calculation
of the covariance matrix in the filter equations (\ref{kxc})

A typical behavior 
%graph
%sizes 
of the magnetic field components~\cite{field} 
as a function of $z$ is shown in fig.~\ref{field}.
The main bending component 
($B_y$) has a bell-shaped $z$--dependence.
The components $B_x,\,B_z$ are sizeable 
%apart 
%outside of
%from 
%the magnet center
away from the central axis
(fig.~\ref{field}b). 
There is clearly no region
%significant 
%subset of the magnet tracking system
%area 
inside the magnet where
%in which t
the field can be regarded as homogeneous.

``Numerical experiments'' with the real track finding procedures
have shown that the approximation~B for derivatives
is accurate enough to be used
%as part of the Kalman filter technique 
for the propagation 
%evaluation
%of the prediction equation  
%{\em prediction equation} ~\cite{Kalman} 
%for 
of the covariance matrix (\ref{ck1})
in the case of the inhomogeneous magnetic field of HERA-B.
In approximation~B the transport matrix has a rather sparse structure:
\begin{equation}
\label{fmk}
F_k = \left( \begin{array}{ccccc}
~~1~~ & ~~0~~ & ~~s_k~~ & ~~0~~ & ~~x'_k~~ \\
0 & 1 & 0 & s_k & 0 \\
0 & 0 & 1 & 0 & t'_k  \\
0 & 0 & 0 & 1 & 0 \\
0 & 0 & 0 & 0 & 1
\end{array}
\right),
\end{equation}
where $s_k=z_k-z_{k-1}$, and derivatives are denoted
$x'_k=\partial x^{k-1}_k / \partial q_{k-1}$ and
$t'_k=\partial t_{xk}^{k-1} / \partial q_{k-1}$
 
For a short distance ({\em $\left|s\right| < 20~{\rm cm}$}) 
the derivatives are obtained 
in a parabolic expansion as:
\[ \partial x^{k-1}_k / \partial q_0 = \frac{1}{2}\, \upsilon\, A_x\, s_k^2
\,\,\,\,{\rm and}\,\,\,\,
\partial t_{xk}^{k-1} / \partial q_0 = \upsilon\, A_x\, s_k\,.\]
For a long distance we find the derivatives 
as the solution of the system (\ref{deriveq}) 
(initial values from (\ref{dq0})),
together with the ``zero trajectory'' (\ref{dxdz}),
by the fourth or fifth-order Runge-Kutta method~\cite{RK}.  

The propagation of the covariance matrix
$F_k C_{k-1} F^T_k$ in (\ref{ck1})
we perform in two steps. First, we define the product of two matrices
$U = F_k C_{k-1}$:  
\begin{equation}
\label{u}
\begin{array}{ll}
u_{11} = c_{11}+c_{13}\,s_k+c_{15}\,x'_k, 
& ~~~~~~~~~~~~~~~~~~~~~~~~~~~~~~~~~~~~~~~~\\
u_{12} = c_{12}+c_{23}\,s_k+c_{25}\,x'_k, & \\ 
u_{13} = c_{13}+c_{33}\,s_k+c_{35}\,x'_k, & \\
u_{14} = c_{14}+c_{34}\,s_k+c_{45}\,x'_k, & \\
u_{15} = c_{15}+c_{35}\,s_k+c_{55}\,x'_k, & \\

u_{21} = c_{12}+c_{14}\,s_k, & \\
u_{22} = c_{22}+c_{24}\,s_k, & \\
u_{23} = c_{23}+c_{34}\,s_k, & \\
u_{24} = c_{24}+c_{44}\,s_k, & \\
u_{25} = c_{25}+c_{45}\,s_k, & \\

u_{31} = c_{13}+c_{15}\,t'_k, & \\  
u_{32} = c_{23}+c_{25}\,t'_k, & \\  
u_{33} = c_{33}+c_{35}\,t'_k, & \\  
u_{34} = c_{34}+c_{45}\,t'_k, & \\  
u_{35} = c_{35}+c_{55}\,t'_k, &  \\ 

u_{4i} = c_{4i}~~~~~& (i=1\div5),  \\
u_{5i} = c_{5i}~~~~~& (i=1\div5),  
\end{array}
\end{equation}
where $c_{ij} = \{C_{k-1}\}_{ij}$
and $u_{ij} = \{U\}_{ij}$.
Then the matrix $U$ is multiplied by $F^T_k$ to obtain the final
symmetric matrix $C' = F_k C_{k-1} F^T_k$:  
\begin{equation} 
\label{uf} 
\begin{array}{ll} 
C'_{11} = u_{11}+u_{13}\,s_k+u_{15}\,x'_k,  
& ~~~~~~~~~~~~~~~~~~~~~~~~~~~~~~~~~~~~~~~~~~~~\\ 
C'_{21} = u_{21}+u_{23}\,s_k+u_{25}\,x'_k,   &              \\
C'_{31} = u_{31}+u_{33}\,s_k+u_{35}\,x'_k,   &              \\
C'_{41} = u_{41}+u_{43}\,s_k+u_{45}\,x'_k,   &              \\
C'_{51} = u_{51}+u_{53}\,s_k+u_{55}\,x'_k,   &              \\

C'_{22} = u_{22}+u_{24}\,s_k,   &              \\ 
C'_{32} = u_{32}+u_{34}\,s_k,   &              \\ 
C'_{42} = u_{42}+u_{44}\,s_k,   &              \\ 
C'_{52} = u_{52}+u_{54}\,s_k,   &              \\ 

C'_{33} = u_{33}+u_{35}\,t'_k,   &              \\ 
C'_{43} = u_{43}+u_{45}\,t'_k,   &              \\ 
C'_{53} = u_{53}+u_{55}\,t'_k,   &              \\ 

C'_{44} = u_{44}, & \\
C'_{54} = u_{54}, & \\ 
C'_{55} = u_{55}. & \\ 

\end{array} 
\end{equation} 
%where $C'_{ij} = \{F_k C_{k-1} F^T_k\}_{ij}.$
The evaluation of the elements in (\ref{u}) and (\ref{uf})
%includes 
implies 37 multiplications, that is much smaller than the 200
multiplications needed for the case of the completely filled matrix $F_k$.   

The product of the matrices $K_k$ and $H^{(1)}_k$ in (\ref{kxc}) is 
\begin{equation} 
\label{kh1} 
K_k\,H^{(1)}_k = \left( \begin{array}{ccccc} 
~k_1\,h_1~~ & ~~0~~ & ~~0~~ & ~~0~~ & ~~0~~ \\ 
~k_2\,h_1~~ & ~~0~~ & ~~0~~ & ~~0~~ & ~~0~~ \\ 
~k_3\,h_1~~ & ~~0~~ & ~~0~~ & ~~0~~ & ~~0~~ \\ 
~k_4\,h_1~~ & ~~0~~ & ~~0~~ & ~~0~~ & ~~0~~ \\ 
~k_5\,h_1~~ & ~~0~~ & ~~0~~ & ~~0~~ & ~~0~~ \\ 
\end{array} 
\right), 
\end{equation} 
and for the matrix $H^{(2)}$ 
\begin{equation}  
\label{kh2}  
K_k\,H^{(2)}_k = \left( \begin{array}{ccccc}  
~k_1\,h_1~~ & ~~k_1\,h_2~~ & ~~0~~ & ~~0~~ & ~~0~~ \\  
~k_2\,h_1~~ & ~~k_2\,h_2~~ & ~~0~~ & ~~0~~ & ~~0~~ \\  
~k_3\,h_1~~ & ~~k_3\,h_2~~ & ~~0~~ & ~~0~~ & ~~0~~ \\  
~k_4\,h_1~~ & ~~k_4\,h_2~~ & ~~0~~ & ~~0~~ & ~~0~~ \\  
~k_5\,h_1~~ & ~~k_5\,h_2~~ & ~~0~~ & ~~0~~ & ~~0~~ \\  
\end{array}  
\right).  
\end{equation}  
For the case with the matrix $H^{(2)}$ the covariance matrix $C_k$ 
is given by:
\begin{equation}  
\label{ck}  
\begin{array}{lll}  
\{C_k\}_{11} = & C_{11} - k_1\,h_1\,C_{11} - k_1\,h_2\,C_{21}, &
~~~~~~~~~~~~~~~~~~~~~~~~~~~~~~~~\\
\{C_k\}_{12} = & C_{12} - k_1\,h_1\,C_{12} - k_1\,h_2\,C_{22}, & \\ 
\{C_k\}_{13} = & C_{13} - k_1\,h_1\,C_{13} - k_1\,h_2\,C_{23}, & \\ 
\{C_k\}_{14} = & C_{14} - k_1\,h_1\,C_{14} - k_1\,h_2\,C_{24}, & \\ 
\{C_k\}_{15} = & C_{15} - k_1\,h_1\,C_{15} - k_1\,h_2\,C_{25}, & \\ 

\{C_k\}_{22} = & C_{22} - k_2\,h_1\,C_{12} - k_2\,h_2\,C_{22}, & \\  
\{C_k\}_{23} = & C_{23} - k_2\,h_1\,C_{13} - k_2\,h_2\,C_{23}, & \\  
\{C_k\}_{24} = & C_{24} - k_2\,h_1\,C_{14} -  k_2\,h_2\,C_{24}, & \\  
\{C_k\}_{25} = & C_{25} - k_2\,h_1\,C_{15} - k_2\,h_2\,C_{25}, & \\  

\{C_k\}_{33} = & C_{33} - k_3\,h_1\,C_{13} - k_3\,h_2\,C_{23}, & \\   
\{C_k\}_{34} = & C_{34} - k_3\,h_1\,C_{14} - k_3\,h_2\,C_{24}, & \\   
\{C_k\}_{35} = & C_{35} - k_3\,h_1\,C_{15} - k_3\,h_2\,C_{25}, & \\   

\{C_k\}_{44} = & C_{44} - k_4\,h_1\,C_{14} - k_4\,h_2\,C_{24}, & \\    
\{C_k\}_{45} = & C_{45} - k_4\,h_1\,C_{15} - k_4\,h_2\,C_{25}, & \\    

\{C_k\}_{55} = & C_{55} - k_5\,h_1\,C_{15} - k_5\,h_2\,C_{25}. & \\     

\end{array}      
\end{equation}   
Only 40 multiplications are sufficient
%is enough 
to obtain the matrix $C_k$ in this case and  
20 multiplications for the option with the matrix $H^{(1)}$.   
This has to be compared with  
100 multiplications needed for the completely filled matrix $H$.    
The optimized matrix evaluation was
implemented in functions of the {\em ranger } 
package related to the magnet tracking~\cite{ranger}. 

\section{Program implementation}

The described approach  for optimized integration of
the equations~\ref{dxdz} and their derivatives 
was implemented as a set of C++ functions 
included in the {\em ranger } package
and used for the magnet tracking~\cite{ranger} and the track 
fit~\cite{refit, momres}.
All functions are of type {\tt void}.
In the following , we list the definition of common parameters of
these functions in C++:
\begin{verbatim}
//           
//           INPUT PARAMETERS
//
  double z_in ;    // z value for input parameters
  double p_in[5];  // vector of input track parameter (x,y,tx,ty,Q/p)
  double c_in[25]; // covariance matrix of input parameters
  float  error[2]; // desired accuracy in cm
                        // error[0] for Inner Tracker region
                        // error[1] for Outer Tracker region
//
//           OUTPUT PARAMETERS
//
  double z_out;    // z value for output parameters
  double p_out[5]; // vector of output track parameters
  double rkd[25];  // derivatives of output parameters with respect 
                   // to input
                   // rkd[0] deriv. of p_in[0] with respect to p_out[0]
                   // rkd[1]           p_in[0]                 p_out[1]
                   //                    . . .
                   // rkd[5]           p_in[1]                 p_out[0]
                   //                    . . .
                   // rkd[24]          p_in[4]                 p_out[4]

  double c_out[25];// covariance matrix of output parameters
  int    ierror;   // error flag ( = 0 ok, = 1 particle curls)
\end{verbatim}

%\newpage
The definition of the corresponding variables in FORTRAN is~:
\begin{verbatim}
c           
c           INPUT PARAMETERS
c
      real*8 z_in       ! z value for input parameters
      real*8 p_in(5)    ! vector of input parameters (x,y,tx,ty,Q/p)
      real*8 c_in(5,5)  ! covariance matrix of input parameters
      real  error(2)    ! desired accuracy in cm
c                           error(1) for Inner Tracker region
c                           error(2) for Outer Tracker region
c
c           OUTPUT PARAMETERS
c
      real*8 z_out      ! z value for output parameters
      real*8 p_out(5)   ! vector of output track parameters 
      real*8 rkd(5,5)   ! rkd(i,j) derivative of p_in(i) with respect
c                         to p_out(j)
      real*8 c_out(5,5) ! covariance matrix of output parameters
      integer ierror    ! error flag ( = 0 ok , = 1 particle curls)
\end{verbatim}
A typical function call in C++ 
\begin{verbatim}
  rk4order_(double& z_in, double* p_in,double& z_out, double* p_out,
            double* rkd, int& ierror);
\end{verbatim}
invokes function integrating equations for particle parameters 
and equations for derivatives in the approximation~A
by a fourth-order Runge-Kutta method . 

In FORTRAN this function can be invoked like {\tt subroutine}~:
\begin{verbatim}
      call rk4order(z_in, p_in, z_out, p_out, rkd, ierror)
\end{verbatim}
The differences are evident and in following, the call statements in
FORTRAN will be mentioned only.

A statement 
\begin{verbatim}
      call rk4fast(z_in, p_in, z_out, p_out, rkd, ierror)
\end{verbatim}
effects integration of the equations for the `zero trajectory'
and calculation of derivatives in the approximation~B 
by a fourth-order Runge-Kutta method . 

The function called as :
\begin{verbatim}
      call rk1fast(z_in, p_in, z_out, p_out, rkd, ierror)
\end{verbatim}
calculates parameters and derivatives in approximation~B using
a parabolic expansion of the particle trajectory.

%\newpage
The function evaluating particle parameters (approximation~A
for derivatives) by a fifth-order
Runge-Kutta method with adaptive step size control is invoked 
by the statement~:

\begin{verbatim}
      call rk5order(z_in, p_in, error, z_out, p_out, rkd, ierror)
\end{verbatim}
The corresponding function which evaluates the derivatives in
approximation~B
is executed by :
\begin{verbatim}
      call rk5fast(z_in, p_in, error, z_out, p_out, rkd, ierror)
\end{verbatim}
The function invoked as: 

\begin{verbatim}
      call rk5numde(z_in, p_in, error, z_out, p_out, rkd, ierror)
\end{verbatim}
evaluates  the output parameters by a fifth-order
Runge-Kutta method with adaptive step size control
and calculates derivatives by `numerical differentiation' of 
the output parameters as a function of the input parameters. The field
gradients are not neglected in this case but the function spends by factor
of 5 more computing time than {\tt rk5fast}.

The ``fully automatic'' function {\tt rktrans} transports particle
parameters from plane $z_{0}$ to plane $z$ 
 by means of three different methods depending on 
$h=z-z_{0}$ as it was described earlier. 
The derivatives are calculated in approximation~B.
The call statement for this function is :

\begin{verbatim}
      call rktrans(z_in, p_in, z_out, p_out, rkd, ierror)
\end{verbatim}
The function {\tt rktransc } uses a similar
approach to transport particle parameters and the corresponding
covariance matrix . 
The structure of the derivative matrix in approximation~B is fully 
exploited for maximum speed.
The function can be invoked by a statement 

\begin{verbatim}
      call rktransc(z_in, p_in, c_in, z_out, p_out, c_out, ierror)
\end{verbatim}
All described functions were designed for the conditions
of the complete geometry of the HERA-B detector where typical transport
distances  are $50 - 100\;{\rm cm}$. The limitations of tracking precision
coming from multiple 
scattering and measurement errors in trackers constrain the tracing 
accuracy required for such distances.

The function {\tt rk5clip} can be
used to propagate particle parameters over larger distances with higher
accuracy (approximation A for derivatives).
The call statement for this function is

\begin{verbatim}
      call rk5clip(z_in, p_in, z_out, p_out, rkd, ierror)
\end{verbatim}
The following procedure was used for the tuning of the steering parameters. 
Particle parameters were generated with production slopes  $t_{x,y}$
uniformly distributed 
from $-100\; {\rm mrad}$ to $100\; {\rm mrad}$. Each particle was 
traced by small
steps from the target to the area behind the magnet
$(z=700\; {\rm cm})$ and then traced back using {\tt rk5clip}. 
The difference between initial and final particle coordinates 
was regarded as a measure of computational accuracy. 
The accuracy for selected steering parameters is shown 
in the table.
As expected, the function {\tt rk5clip} spends more computing time especially
for low momentum 
(the dependence is roughly $\propto 1/p_{0}$)~.
\begin{table}
\begin{center}
\begin{tabular}{|c|cc|}
\hline \hline
$p_{0}$ & $rms(x)$ &  $rms(y)$  \\   
${\rm GeV/c}$  & ${\rm \mu m}$ & ${\rm \mu m}$  \\ \hline \hline
$5$  & $12$ & $13$  \\
$10$ & $ 9$ & $6$ \\
$30$ & $ 9$ & $6$ \\
$60$ & $ 7$ & $6$ \\
$90$ & $ 7$ & $3$ \\  \hline
\end{tabular}
\end{center}
\caption{The computational accuracy of {\tt rk5clip} for different
momenta}
\label{tab:accmag}
\end{table}
It should be noted that all described
functions do not check if the magnetic field is defined in the
region where the particle should be traced. 
The user himself must make sure that a corresponding function is
invoked within the 
magnetic field and should use linear line propagation in the field-free case. 
Also, an attempt to trace a very slow particle will not be successful
because the equations (1) do not describe a particle curling in 
the magnetic field. 

An additional function, called as :
\begin{verbatim}
      real zmin,zmax   !  in cm
         . . .
      call rkzfield(zmin,zmax)
\end{verbatim}
returns as output the lower and upper bounds {\tt zmin, zmax}  (with
respect to the center of the magnet) of the region where the field is defined 
by routines {\tt gufld, utfeld}~\cite{field}.

This region ( roughly $ \pm 5~{\rm m}$ from the center of the magnet)
includes the area where we have field measurements 
or at least results from the MAFIA calculation.
At the edges of the region the magnetic field can be neglected so that simple
linear line searches and fits are sufficient for pattern recognition
and track fitting ~\cite{Lohse}. This can save computing time for
event generation and reconstruction.
Therefore during the event simulation in HBGEAN ~\cite{HBGEAN} 
the procedure for
particle tracing in the magnetic field is invoked only when the particle 
is within the GEANT volume {\tt MAGN} . 
The definition of this volume can be
found in the {\em Arte} table {\tt GESL} for the detector component 
{\tt Magnet}. 
Note that this volume is not identical with the region of non-zero field
as defined by {\tt gufld, utfeld}.

\section*{Acknowledgments}
I am greatly indebted to Rainer Mankel, the author of the {\em ranger}
package,  for sharing with me his
experience in the field of track finding and knowledge about {\em ranger}. 
I would like to thank him for the careful reading of the manuscript
and the useful advices.
I am indebted to G.\,Bohm and H.\,Kolanoski for the discussion
of the revised version.
I am very much grateful to DESY Zeuthen for the kind 
hospitality during my visit.

%\clearpage

\end{document}